# A Rayleigh Bénard Convective Instability Study Using Energy Conserving Dissipative Particle Dynamics

Anuj Chaudhri [a] & Jennifer R. Lukes [b]

Department of Mechanical Engineering and Applied Mechanics,

University of Pennsylvania, Philadelphia, PA 19104 USA

A Rayleigh Bénard instability study using the energy conserving dissipative particle dynamics method is presented here for the first time. The simulation is performed on an ideal dissipative particle dynamics fluid in a three dimensional domain with carefully selected parameters to make the convection terms in the equation more dominant than the conduction ones. Beyond a critical temperature difference a two cell pattern is observed as the dominant structure. As the temperature is increased further, the density changes in the system are sharp with formation of distinct high density layers close to the cold wall. Doubling the length of the domain led to the formation of four convection cells with the same cell diameter as before, confirming the invariance of the pattern formation in that dimension. Changes in the height of the domain led to cells with more uniform looking patterns. The results and patterns seen here are qualitatively similar to previous studies performed on rarefied gases.

[a] Presently at University of Chicago, Chicago IL 60637,   Email: chaudhri@uchicago.edu

[b] Email: jrlukes@seas.upenn.edu

# I. INTRODUCTION

Systems that are driven away from equilibrium give rise to spatio-temporal patterns also known as dissipative structures [1,2]. These patterns have been observed in fluidic, biological, and chemical systems and in systems undergoing phase transformations [3,4]. A large focus of nonequilibrium physics is the development of stability theories to help understand these systems better. In the case of fluids, hydrodynamic instability is a well-studied problem [5]. The onset of convective patterns due to surface tension variations with temperature was studied experimentally by Bénard [6]. The solution to the problem of convection cells due to density variations with temperature was first proposed by Lord Rayleigh [7]. For historical reasons, the thermal instability of a fluid layer heated from below is called Rayleigh-Bénard convection.

Rayleigh-Bénard convection has been widely studied both experimentally and theoretically [8,9,10]. It is the simplest problem that involves the formation of stable convection cell patterns in a layer of fluid heated from below. An external field acts adversely to the imposed temperature gradient. The heated fluid being 'top-heavy' rises against the external field and cold fluid rushes to take its place. This generates a density gradient in the system and sets up convection rolls that become stable over time. This can be treated using linear stability analysis under the Boussinesq approximation to the Navier-Stokes equations. A very simple and clear treatment of the analysis can be found in [11].

The transition to convection, as predicted by linear stability analysis, for a mechanically incompressible fluid is governed by a dimensionless quantity known as the Rayleigh number (Ra).

$$\mathrm{Ra} = \frac{g\beta k_B \Delta T H^3}{\alpha v} \quad (1)$$

Here $g$ is the strength of the external force, $\beta = 1/k_B T$ is the volume expansion coefficient for the ideal fluid, $\Delta T$ is the difference in temperatures between the hot and cold walls, $H$ is the height of the domain, $\alpha$ is the thermal diffusivity and $v$ is the kinematic shear viscosity. Ra represents a balance between the buoyancy forces in the system to other forces that arise from momentum and thermal diffusion. When the Ra is below a critical value, heat transfer in the system is primarily due to conduction and when it exceeds the critical value, heat transfer is dominated by convection effects. In the Rayleigh-Bénard problem, instability occurs at the minimum temperature gradient at which the kinetic energy dissipated by viscosity balances the internal energy released by the buoyancy forces [5]. The critical Ra gives a quantitative estimate of this phenomenon.

The problem of observing convective cells has also attracted interest by researchers using nonequilibrium molecular dynamics (MD) simulations [12,13,14,15,16,17,18]. The first few MD studies focused on qualitative comparisons of MD simulations with the linear stability results [12-14]. The study by Mareschal et al. [12,13] was performed for two-dimensional non-periodic hard disk systems for millions of time steps whereas the study by Rapaport [14] was done in two-dimensional periodic hard disk systems. In the case of hard disks, the transport coefficients are available using the Enskog solution to the Boltzmann equation.

Mareschal et al. assumed that the transport properties were constant over the temperature differences in their simulations, which enabled them to estimate Ra for their systems. On the other hand, Rapaport did not calculate Ra because of the strong variation of local transport coefficients arising from convective patterns, system size, dimensionality and changes in temperature across his system. Both studies reported formation of stable convective patterns with multiple rolls. The studies confirmed that MD simulations require very large temperature gradients and external fields to overcome the thermal noise present in small systems. Rapaport [18] also investigated time dependent convective cell patterns using MD. Higher Rayleigh number simulations were performed to show evidence of temporal periodicity in two-dimensional hard-disk systems.

The other studies were focused on quantitative comparisons of MD simulations with continuum equations [15, 17]. These studies were performed on systems of hard disks in two-dimensional non-periodic boxes for which the transport coefficients are well known from kinetic theory. The studies found that at small ratios of temperature gradient to average temperature, the Boussinesq approximation works well in predicting the critical Ra. They also found that the aspect ratio plays an important role in determining whether single or double roll cells would appear in the system. The authors in [16] also performed their studies in two-dimensional non-periodic systems and concluded that velocity boundary conditions at the hot and cold walls played a vital role in producing the convection patterns. In contrast to the previous authors, they argue that there is no consensus on the validity of the Boussinesq approximation in MD simulations.

Of late, mesoscopic approaches have become very important in simulating instabilities in fluids. Recent studies have looked at the Rayleigh-Bénard problem using smooth particle applied mechanics [19] and lattice Boltzmann techniques [20]. In this paper we study the Rayleigh-Bénard convection problem using the energy conserving version of dissipative particle dynamics (DPD) for the first time. We begin by reviewing the basic concepts of energy conserving DPD in section II. The details of the method and computational setup are also given in section II. Section III describes the simulation results for the systems described in section II. Section IV concludes with a summary of the key results of the paper.

## II. METHODOLOGY

**REVIEW OF ENERGY CONSERVING DPD**

DPD is a meshless, coarse-grained, particle based method used to simulate systems at mesoscopic length and time scales. In DPD, the solvent is modeled explicitly and includes the hydrodynamic interactions. Essentially atoms, molecules or monomers (denoted as 'particles') are grouped together into mesoscopic clusters (denoted 'beads') that are acted on by conservative, dissipative and random forces. The interaction forces are pairwise in nature and act between bead centers. The theoretical basis for DPD has been studied extensively and the equilibrium basis for isothermal situations has been identified [21]. The link between the theoretical development and application of the method has been established subsequently [22, 23]. The original formulation of DPD conserves linear and angular momentum but not energy [24]. To model heat flow in mesoscopic systems using DPD, an additional internal energy variable is introduced [25, 26]. The internal energy variable represents the relaxed degrees of freedom of the particles that make up the bead. The mechanical energy dissipated due to velocity dependent forces is transformed into the internal energy of the beads. In addition to this a model of heat conduction is also introduced to account for heat transfer processes that take place due to changes in internal energy. Since beads can have different internal energies, temperature gradients can easily be modeled. This model has been used to study one-dimensional heat conduction in a random "frozen" DPD solid [27, 28, 29], two-dimensional heat conduction [29], and convection rolls in simple square cells [30].

The DPD governing equations are stochastic ordinary differential equations (SODE) that represent continuous Markov processes of the Langevin type. The dimensionless equations of motion for the $i^{th}$ bead with position coordinate $\bar{\mathbf{r}}_i$, velocity coordinate $\bar{\mathbf{v}}_i$, and mass $\bar{m}$ are given by:

$$\frac{d}{d\bar{t}}\ \bar{\mathbf{r}}_i\ =\ \bar{\mathbf{v}}_i \tag{2}$$

$$\bar{m}\frac{d}{d\bar{t}}\ \bar{\mathbf{v}}_i\ =\ \bar{\mathbf{F}}_i^{\mathrm{C}} + \bar{\mathbf{F}}_i^{\mathrm{D}} + \bar{\mathbf{F}}_i^{\mathrm{R}} \tag{3}$$

where $\bar{\mathbf{F}}_i^{\mathrm{C}}$, $\bar{\mathbf{F}}_i^{\mathrm{D}}$, and $\bar{\mathbf{F}}_i^{\mathrm{R}}$ are the conservative, dissipative and random forces acting on the $i^{th}$ bead where the overbar denotes parameters in dimensionless DPD units. The energy-conserving model includes an additional equation that needs to be solved for the internal energy variable $\bar{\varepsilon}_i$ for the $i^{th}$ bead:

$$d\,\bar{\varepsilon}_i \;=\; d\!\left[\bar{\varepsilon}_i^{\text{VH}}\right] + d\!\left[\bar{\varepsilon}_i^{\text{HC}}\right] \tag{4}$$

where the first term on the right hand side in Eq. (4) arises from changes in mechanical energy of the system that lead to viscous heating. The second term arises from differences in internal energy content between the beads. This bead-bead energy transfer is modeled as a mesoscopic heat conduction (HC) term. More details on the functional forms of the force and energy terms used in Eqs. (3) and (4) can be found in these references [22, 26, 29, 31]. For the Rayleigh-Bénard system the entire set of equations (2)-(4) need to be solved. The computational setup and details of the parameters used are described in the next section. All parameters for the simulations are given in Table 1. All the simulations were performed using the simple weight function and the Groot-Warren integrator described in reference [22].

**COMPUTATIONAL SETUP AND SYSTEM PARAMETERS**

The Rayleigh-Bénard system of DPD beads is configured in a rectangular domain as shown in Fig. 1. Three different aspect ratios $\bar{\delta} = \bar{L}/\bar{H}$ of 5.56, 2.78 and 1.79 were studied in this paper. The value of the cell diameter $\bar{d}$ is estimated after a stable pattern has developed and is compared for the different cases studied here. The system is periodic in the two (X and Z) directions with 'walls' in the third (Y) direction. The bottom and top walls serve as the 'source' and 'sink' of energy to the system. In DPD, the general convention is to non-dimensionalize the equations of motion and all the variables associated with the system. More details can be found in [29, 31]. Following that convention, the thermal energy of the system is used interchangeably with the temperature, as they only differ by a constant $k_B$. The bottom wall is kept at a higher temperature $\overline{k_B T_{\text{hot}}}$ than the top wall, which is at a lower temperature $\overline{k_B T_{\text{cold}}}$. This sets up a temperature gradient in the system. For beads that cross the walls at the boundaries, the bead positions $\overline{\mathbf{r}}_i$ are specularly reflected, the normal velocity component is reversed, and the tangential velocity component is left unchanged. This would be equivalent to applying full-slip boundary conditions at the continuum level. The temperature boundary condition is enforced by fixing the temperature of beads that cross the walls to the wall temperature.

An external field $\overline{g}$ is applied to the beads in the negative y-direction. In MD Rayleigh-Bénard studies, the strength of the external force $\overline{g}$ is often guided by the ratio $\overline{R}$ of the potential energy difference across the domain to the particle kinetic energy difference across the domain. In such studies, $\overline{R}$ rather than $\mathrm{Ra}$ is often used to indicate the strength of the Rayleigh-Bénard convection because $\overline{R}$ is more convenient to compute than $\mathrm{Ra}$ and also has a similar physical interpretation. This convention is also followed in the present DPD study. Since particle energy in MD is related through equipartition to temperature, $\overline{R}$ can be expressed as [13, 16].

$$\overline{R} = \frac{\overline{m}\overline{g}\overline{H}}{k_B \Delta T} \tag{5}$$

Here $\overline{m}$ is the mass of a particle and $\overline{H}$ is the height of the system. Typically $\overline{R}$ in MD studies is chosen to be one and the value of $\overline{g}$ is calculated based on the system parameters. In contrast, $\bar{g}$ is fixed in the present DPD simulations and the value of $\bar{R}$ is calculated. $\bar{R}$ is varied in the present simulations by changing the value of $\overline{k_B \Delta T}$ such that the average temperature remains constant. The temperatures used in the simulations are based on the bead internal energy. By fixing the average temperature in the runs, it is assumed that the average properties of the system such as the volumetric thermal expansion coefficient, shear viscosity and thermal diffusivity do not change and that the different cases where $\overline{k_B \Delta T}$ is changed can be compared. The variables and the resulting calculated values used in this chapter are shown in Table 2.

Mackie et al. [30] investigated the formation of convection cells in closed square cells with the temperature gradient orthogonal to the gravitational field. The notation and nondimensionalization presented in their study are different from what was presented by us earlier [29]. However, we follow their analysis here to make our own calculations to determine the parameters needed for the present convective studies. The main parameters that need to be chosen for the simulations are the strength of the noise $\bar{\sigma}$ and strength of the heat conduction $\bar{K}_o$ parameter (see Table 1). The friction parameter $\bar{\gamma}$ and random heat flux parameter $\bar{\Lambda}^\varepsilon$ can be set using the equations shown in Table 1. The convection cells are formed when the energy transfer due to momentum relaxes faster than the energy transfer due to heat conduction. The result is a convection dominated regime as opposed to a conduction dominated one. The ratio of the time scales associated with energy and momentum transfer is calculated to be 2.2e-3 in close agreement with Mackie et al. (1999).

For the parameters listed in Table 1, the shear viscosity and diffusion constant can be calculated using the analytical expressions given by Marsh et al. (1997). The Schmidt number can be computed from these and was found to be $\sim 1$, hence the DPD fluid modeled here represents a gas. The conservative force parameter $\bar{a}$ is also set to zero (Table 1), which forces the pressure equation of state to represent an ideal gas. Rayleigh-Bénard convection has been studied in compressible rarefied gases for different values of Knudsen and Froude numbers [32] [33] in the literature. The DPD results are compared to these studies wherever applicable.

## III. RESULTS AND DISCUSSION

**CHALLENGES IN RAYLEIGH NUMBER CALCULATION**

As discussed above, Ra is the primary dimensionless parameter used to characterize Rayleigh-Bénard convection. However, in DPD the evaluation of this parameter is not straightforward. Two key issues are 1) calculation of transport properties of the DPD fluid, and 2) applicability of the critical Ra derived from linear stability analysis to DPD systems. Shear viscosity and thermal conductivity are the transport properties needed to calculate Ra. The shear viscosity calculations have been performed by us before in a separate study and we found that the simulation results fall between predictions of standard theories [34]. The thermal conductivity, on the other hand, has not been analyzed either analytically or computationally for the DPD fluid. Only the 'frozen' DPD solid has been studied [27, 28, 29]. A major reason for this is the lack of established heat flux expressions suitable for calculating the heat flux from these moving fluids where the energy and momentum equations are being solved simultaneously. Variability of transport parameters across the domain also complicates matters [14, 18].

Even if appropriate values for both transport parameters could be obtained from DPD, questions remain as to whether the results could be compared to linear stability analysis. The prediction of critical Ra by this analysis assumes the Boussinesq approximation, which ignores the variation of density in the various terms of the Navier-Stokes equations except in the buoyancy term where its effect is more pronounced. The Boussinesq approximation has been questioned for molecular dynamics simulations due to the large gradients that need to be imposed to overcome thermal noise in the system [16]. Whether it holds in our case can only be determined if a full analysis of the nonlinear Navier-Stokes equations is performed using the DPD transport coefficients. The preceding issues require detailed study and are beyond the scope of the present work.

**EFFECT OF TEMPERATURE DIFFERENCE ON PATTERN FORMATION**

Simulations were run at different values of $\overline{k_B \Delta T}$ as shown in Table 2 at constant average temperature. This was done to calculate the critical temperature difference $\overline{k_B \Delta T_{\text{cr}}}$ at which the instability would first appear. For the domain with aspect ratio $\overline{\delta}$ of 2.78 listed in Table 2 (1x10x25, 5000 beads), instability begins to appear at $\overline{k_B \Delta T} = 4$. The results from the simulations have been time averaged for $10^5$ iterations at the end of production runs. Figure 2 compares the time averaged velocity profile for the system at temperature differences of 3, 4, 5 and 10. The cells form briefly but disappear and are not self-sustaining for longer periods of time. At $\overline{k_B \Delta T} = 4$ stable convection patterns first start to appear. Hence this is the critical temperature difference where stable patterns can be seen. The cell structure that is observed is the fundamental mode of the Rayleigh-Bénard instability as predicted by theory. As the temperature difference is increased further to 5 and 10, more uniform flows are observed as shown in Fig. 2. It is important to remember that the domain is periodic in the z-direction and hence the cells can translate freely in that dimension. Another interesting observation is the location of vortex centers as the temperature difference is increased. The vortex centers move closer to each other and closer to the hot (lower) wall with increase in temperature difference. Higher temperature differences create stronger vortex propulsions of the beads from the hot to cold wall.

The time averaged number density is plotted for the $\overline{k_B \Delta T} = 5$ and $\overline{k_B \Delta T} = 10$ cases in Fig. 3. Remarkable differences are observed for the two cases. For the $\overline{k_B \Delta T} = 10$ case strong density gradients indicative of fluid stratification appear near the cold wall. Similar profiles have also been reported previously in the MD studies of hard-disk systems [13, 15-17]. They have attributed this effect to the formation of "molecular boundary layers" next to the walls. These boundary layers can lead to sharper velocity gradients seen in Fig. 2. Also, gradients are large in the center of the domain where the cold fluid is descending. For the $\overline{k_B \Delta T} = 5$ case, the densities are more uniform across the domain, with increases near the hot wall.

The change in vortex position with increasing temperature difference can be explained as follows. Due to continuity, the mass flow rate around the vortex must be constant. The mass flow rate is defined as the product of density, velocity and flow area. The higher velocities in the center of the domains (Figs. 2 and 3) drive the vortices closer to each other because a smaller flow area is needed to maintain a constant flow rate.

The temperature isotherms for the above cases are plotted in Fig. 4. In the $\overline{k_B \Delta T} = 3$ case, the hot and cold fluid layers do not show significant variation in the domain. The contours are predominantly horizontal, as would be expected when heat transfer is purely by conduction. As the temperature difference is increased, the downward dip in the temperature contours becomes more pronounced as shown in Fig. 4. For $\overline{k_B \Delta T} = 10$, the descending cold fluid is concentrated in a thin region close to the middle of the domain compared to the ascending hot fluid that is spread over most of the area of the domain. Also, the isotherms are horizontal and closely spaced at the top of the domain where the ascending hot fluid encounters the cold wall and at the bottom of the domain where the descending cold fluid encounters the hot wall. This occurs because the vertical component of velocity vanishes in these regions, leading to purely conductive heat transport in the vertical direction. The difference in temperature profiles between the $\overline{k_B \Delta T} = 5$ and $\overline{k_B \Delta T} = 10$ cases can be attributed the relative contributions of conduction and convection heat transfer [35]. Fluid convection is more pronounced for $\overline{k_B \Delta T} = 10$, as shown in Fig. 2 where the velocities of the descending cold fluid and ascending hot fluid are higher for that case than for the $\overline{k_B \Delta T} = 5$ case. These higher velocities mean that colder(hotter) fluid is able to travel farther toward the hot(cold) wall before being heated(cooled) by thermal diffusion from(to) that wall. This leads to the sharper temperature gradients in Fig. 4 for $\overline{k_B \Delta T} = 10$.

Since the system is closed, mass continuity must hold in the domain. The descending colder fluid is concentrated in a small area and is denser compared to the rising hot fluid. To find the total mass flow rate around the vortex, the time averaged mass flow flux is calculated at each point along the horizontal line that passes through the center of the vortex. This is done by taking the product of the density and the vertical component of velocity $V_y$ at the horizontal line; the results are plotted in Fig. 5. The total mass flow rates of the descending cold fluid and the ascending hot fluid are then calculated by integrating local mass flux along lines B and A, respectively. These values are reported in Table 3 for the various temperature differences. The total mass flow rates of the cold and hot fluids are very close with some deviations at higher temperature differences. The mass flow rates increase with increasing temperature difference as shown in Table 3.

The cell patterns shown here in Fig. 2 can be qualitatively compared to figures obtained in Rayleigh-Bénard simulations of rarefied gases [32] and show very similar characteristics (for eg. Fig. 4 in [32]). The rarefied gas simulations were done using both direct simulation Monte Carlo and finite difference methods for a span of Knudsen and Froude numbers. Variations of the convection patterns were reported by changing the Knudsen and Froude numbers but by keeping the temperature difference constant. The plots showing velocity and density profiles (Figs. 5, 7 from [34]) are qualitatively very similar to Figs. 2, 7 and 8 (to be discussed in the next section) shown in this paper. This agreement shows that energy conserving DPD is a viable simulation method for the study of Rayleigh-Bénard convection.

**EFFECT OF ASPECT RATIO ON PATTERN FORMATION**

The simulations above were run at an aspect ratio of 2.78. The effect of aspect ratio on pattern formation was investigated by changing the length of the domain and by changing the height. The invariance with changing the length of the domain is a good check because the lateral size of the simulations does not affect the system behavior. For the case where the aspect ratio is 5.56, the velocity and temperature profiles are shown in Fig. 6. For this case four convection cells emerge. The cell diameters for each of these cells are the same as the two cell case. The values of $\bar{R}$ also remain unchanged with changing length and hence should not affect the dynamics in any way.

The maximum, minimum and mean time averaged velocity profiles $V_y, V_z$ are plotted along the Z and Y directions in Figs. 7 and 8. Figure 7 shows the sinusoidal pattern of the instability, which repeats along the length of the domain. The non-zero $V_y, V_z$ velocities close to the walls in Fig. 8 confirm the slip boundary conditions that have been enforced there.

When the height of the domain is changed, the convection patterns are similar to the $\bar{\delta} = 2.78$ case, but formation of the cells occurs at a lower temperature difference (not shown here for brevity). The critical temperature difference in this case was found to be $\overline{k_B \Delta T} = 3$. Since the critical Rayleigh number and hence the critical temperature difference is a constant, changing the height of the domain necessarily leads to a change in the temperature difference if all the other properties are held constant. As the temperature difference is increased beyond the critical value, the patterns become more uniform.

**EFFECT OF AVERAGE TEMPERATURE ON PATTERN FORMATION**

In calculations of the Rayleigh number, the coefficient of volumetric expansion $\overline{\beta}$ is evaluated at the average temperature of the system. Beta is defined through density and temperature variations as [11]:

$$\frac{\delta\overline{\rho}}{\overline{\rho}} = -\overline{\beta}\overline{k_B \Delta T} \tag{6}$$

As discussed above, the Schmidt number is close to 1, so the DPD system is modeled as an ideal gas for which $\overline{\beta} = \dfrac{1}{k_B T_{\text{avg}}}$. Thus, if the average temperature of the system is reduced, the volumetric thermal expansion coefficient increases. For typical gases and liquids, $\beta \sim 10^{-3} \ to \ 10^{-4}$. If a temperature difference in the fluid is of 10°C, then the variation of density can atmost be a few percent (Kundu and Cohen, 2002). Hence for density variations to be small the quantity $\overline{\beta}\overline{k_B \Delta T}$ must be $<< 1$. This is one of the requirements for the Boussinesq approximation to be valid and linear stability theory to be applicable. Since the normalized temperature differences used in the DPD simulations are much higher than that used in continuum studies (Table 2), compressibility effects will be significant. These compressibility effects can cause the density gradients to establish quickly and sharply in the system. This is also confirmed by previous MD studies, which indicate that for compressibility effects to be negligible, the normalized temperature difference $\dfrac{\overline{k_B \Delta T}}{\overline{k_B T_{\text{avg}}}}$ must be much smaller than one.

Increasing the Rayleigh number should produce patterns similar to the cases when the temperature difference between the walls is increased at constant average temperature. A good check of this can be done by changing the average temperature of the system keeping the temperature difference between the plates to be a constant. The effect of average temperature on convection patterns was investigated for the case with aspect ratio of 2.78 and temperature difference of 5 at two different average temperatures of 2.75 and 5.5. The velocity, temperature and density contours for these cases are shown in Fig. 9.

The main difference between the two cases is the vertical location of the vortex centers. The vortex centers for the DPD system with lower average temperature $\overline{k_B T_{\text{avg}}} = 2.75$ are located closer to the hot wall at Y=4.4 compared to Y=5.3 for the other case. In the case with the lower average temperature, $\overline{\beta}$ increases. This means for the same $\Delta \overline{T}$, the density difference between the hot and cold fluid is more pronounced. By mass conservation, the higher density of cold fluid and lower density of hot fluid for the case of lower average temperature results in a reduction of flow area for cooler fluid and an increase of flow area for hotter fluid.

It is interesting to compare the velocity, temperature and density profiles for $\overline{k_B \Delta T} = 5$ ; $\overline{k_B T_{\text{avg}}} = 2.75$ in Fig. 9 with the case $\overline{k_B \Delta T} = 10$ ; $\overline{k_B T_{\text{avg}}} = 5.5$ in Figs. 2, 3 and 4. Both these cases have the same change in density to density values, which from Eq. (6) can be written as $\frac{\delta \overline{\rho}}{\overline{\rho}} = \frac{10}{5.5} = \frac{5}{2.75}$. This indicates that the density maps should be quantitatively the same for both the cases and that there should be qualitative similarities between the velocity and temperature profiles. However, the velocity profile for $\overline{k_B \Delta T} = 5$ ; $\overline{k_B T_{\text{avg}}} = 2.75$ in Fig. 9 is 'flatter' compared to that of $\overline{k_B \Delta T} = 10$ ; $\overline{k_B T_{\text{avg}}} = 5.5$ in Fig. 2. Also, the temperature profile in Fig. 9 shows strongly curved isotherms near the left and right edges of the domain (ascending fluid region) while the corresponding profile in Fig. 4 has much flatter isotherms there. In addition, Fig. 9 shows higher density in the lower portions of the ascending fluid region (near the hot wall) than the corresponding density profile in Fig. 3. These differences can be attributed to the change in thermal conductivity as average temperature changes. For an ideal gas, the thermal conductivity is proportional to the square root of temperature. As the average temperature is reduced, the thermal conduction is reduced relative to the convection.

The strong curvature of isotherms in the ascending fluid region of Fig. 9 reflects the increased role of convection: the hot fluid rises more quickly than its thermal energy can be conducted to colder fluid, hence the deviation from horizontal isotherms. The density profile in Fig. 9 tells a similar story. The densities are higher in the ascending fluid region than for the corresponding case in Fig. 3 because the convection is more effective at pulling the colder, denser fluid down to the wall, outward along the wall, and upward. The velocity profiles in Fig. 9 are a reflection of the density profiles. As the fluid descends and then turns to flow horizontally along the hot wall, its density decreases less dramatically than in Fig. 3. The smaller decrease in density leads to a smaller increase in flow area around the vortex and a flatter velocity profile.

## IV. CONCLUSIONS

A Rayleigh-Bénard convective instability study was performed using the energy conserving version of dissipative particle dynamics for the first time. The fundamental mode of the instability (two cell pattern) is observed at a critical temperature difference for domains with aspect ratios of 2.78. At higher temperature differences, the cold fluid is concentrated in a thin region close to the middle of the cell. There is also considerable stratification of fluid as the temperature difference is increased. By changing the length of the domain to a larger aspect ratio of 5.56, the four cell pattern is found to be the dominant one. However the cell diameter is found to be the same as that of the two cell pattern. When the height of the domain is changed the two cell pattern is found at a lower critical temperature. As the average temperature of the system is changed keeping the temperature difference constant, the cell patterns are found to be similar to the case when the temperature difference is changed at constant average temperature. These effects are attributed to variations of density and compressibility with temperature.

**Fig. 1** Schematic showing the Rayleigh-Bénard computational setup

**Fig. 2** a) Time averaged velocity contours for system with Aspect Ratio = 2.78, N = 5000 at temperature differences of 3, 4, 5 and 10, and b) Time averaged velocity $V_y$ at temperature differences of 5 and 10

**Fig. 3** Time averaged number density plot for system with Aspect Ratio = 2.78, N = 5000 at temperature differences of 5 and 10

**Fig. 4** Time averaged isotherms for system with Aspect Ratio = 2.78, N = 5000 at temperature differences of 3, 4, 5 and 10

**Fig. 5** Schematic showing the lines A and B for the total mass flow rate calculations; Time averaged mass flux for system with Aspect Ratio = 2.78, N = 5000 at temperature differences of 4, 5 and 10

**Fig. 6** Time averaged velocity contours and isotherms for system with Aspect Ratio = 5.56, N = 10000 at temperature differences of 5 and 10

**Fig. 7** Maximum, minimum and mean $V_y$ and $V_z$ velocity variations along the Z – direction for system with N = 10000, averaged along the x-y plane

**Fig. 8** Maximum, minimum and mean $V_y$ and $V_z$ velocity variations along the Y – direction for system with N = 10000, averaged along the x-z plane

**Fig. 9** Time averaged velocity contours, isotherms and number density for system with Aspect Ratio = 2.78, N = 5000, delta T = 5 at average temperatures of 5.5 and 2.75

**Table 1 Values of dimensionless DPD parameters (in DPD units) used in the Rayleigh-Bénard simulations**

**Table 2 Values of system variables used in the Rayleigh-Bénard simulations**

**Table 3 Total mass flow rate variation with temperature difference for system with N = 5000, Aspect Ratio = 2.78**

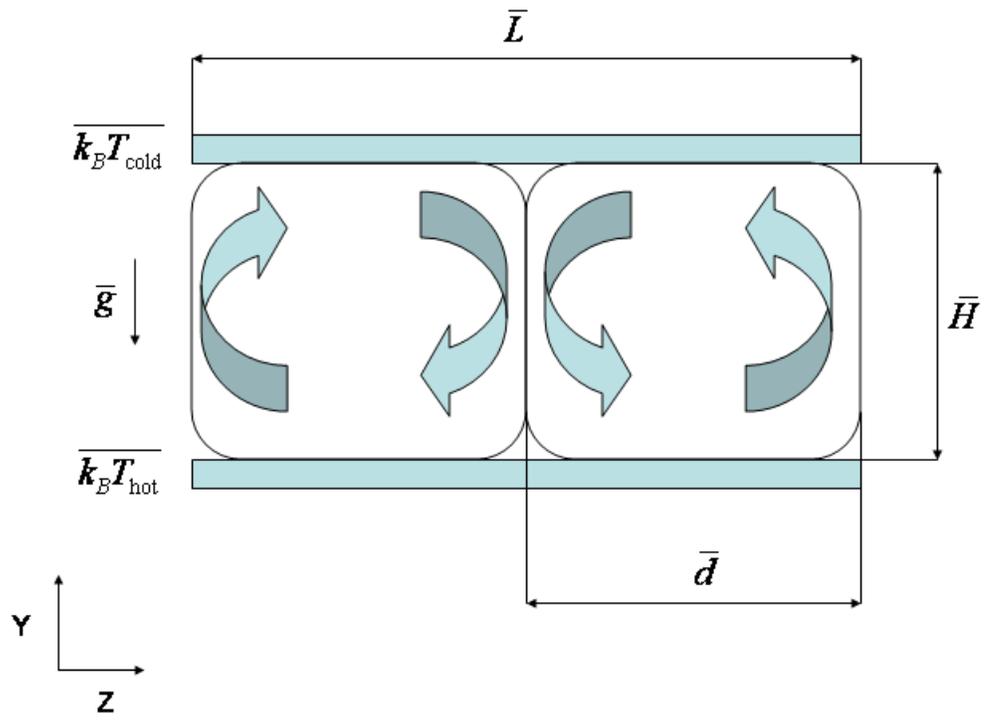

**Fig. 1**

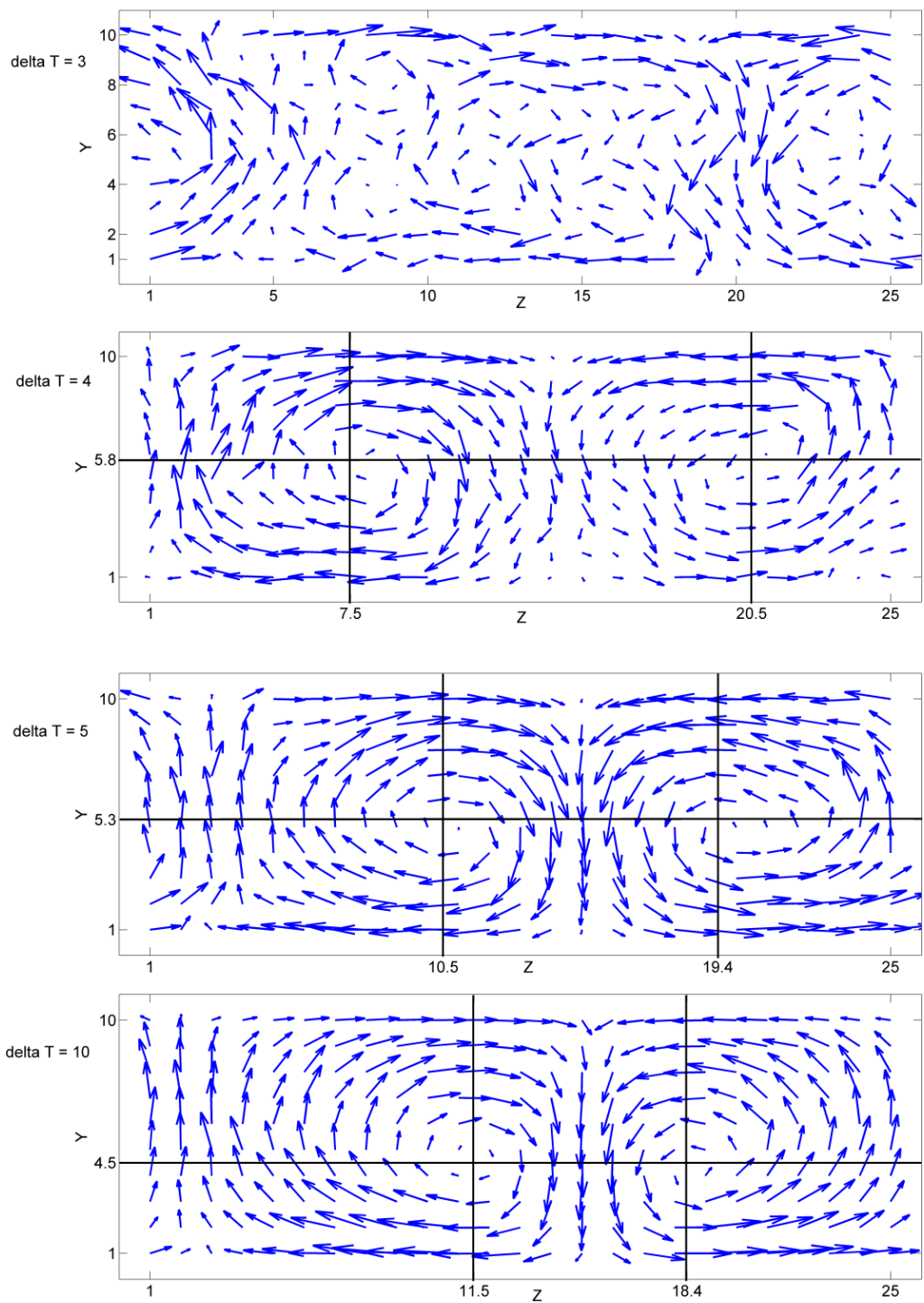

**Fig. 2a**

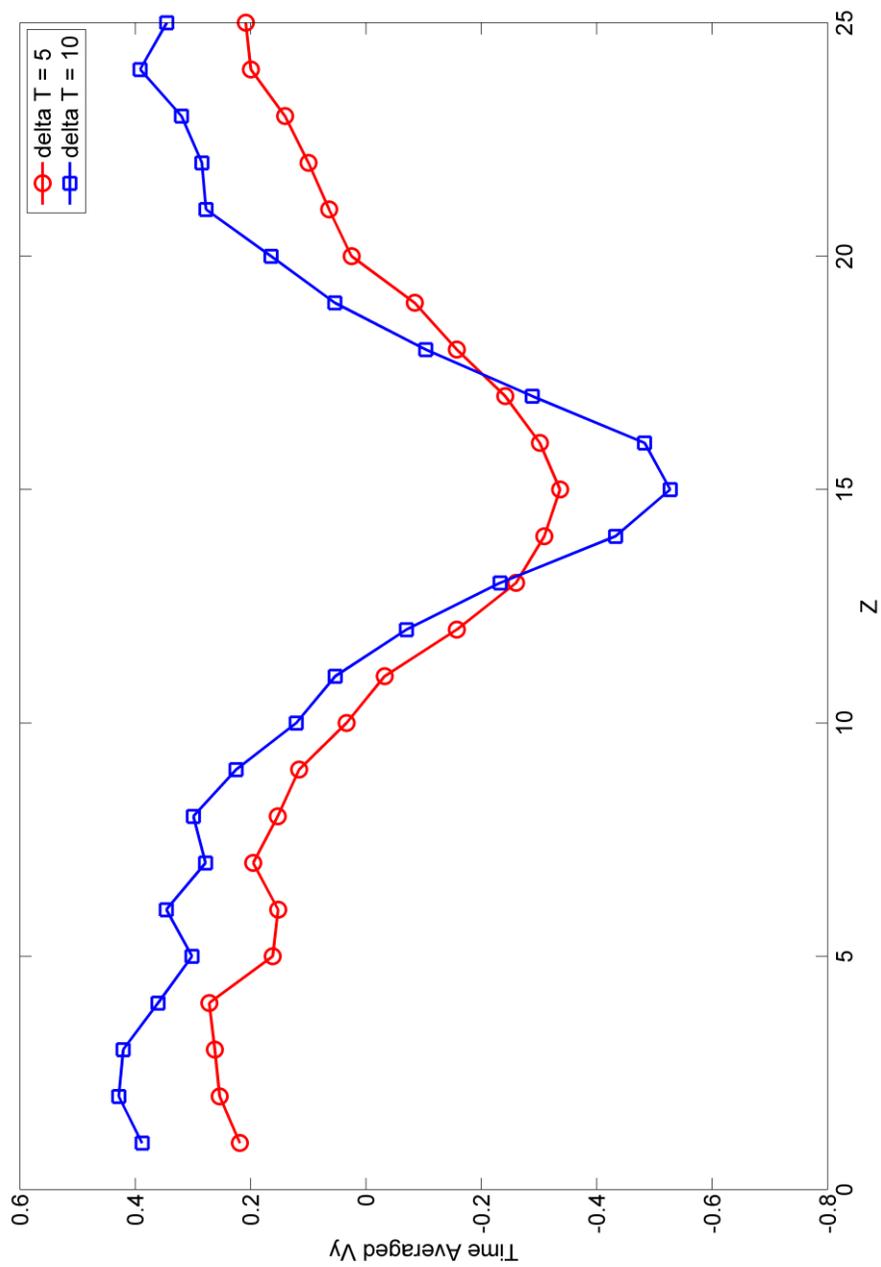

**Fig. 2b**

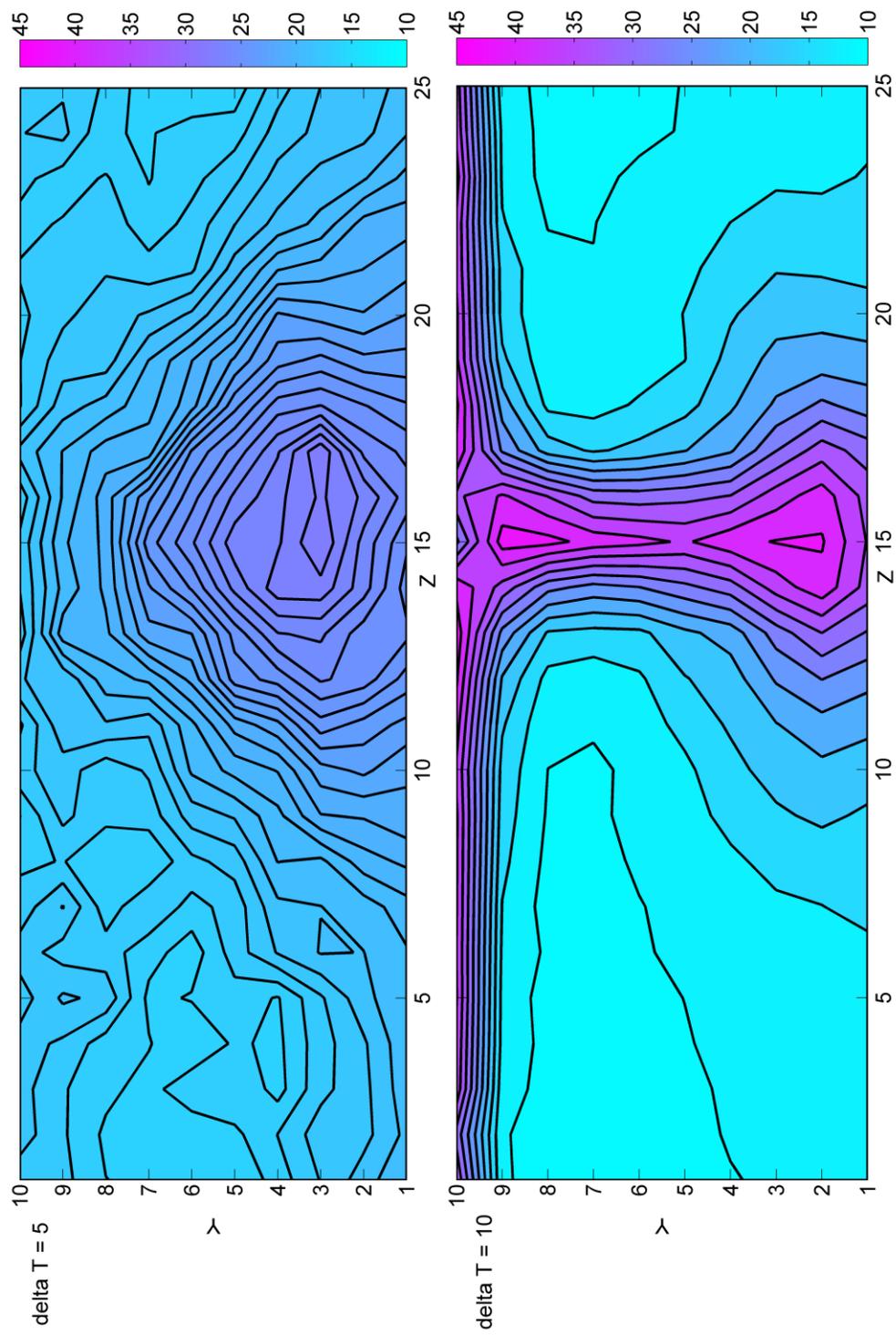

**Fig. 3**

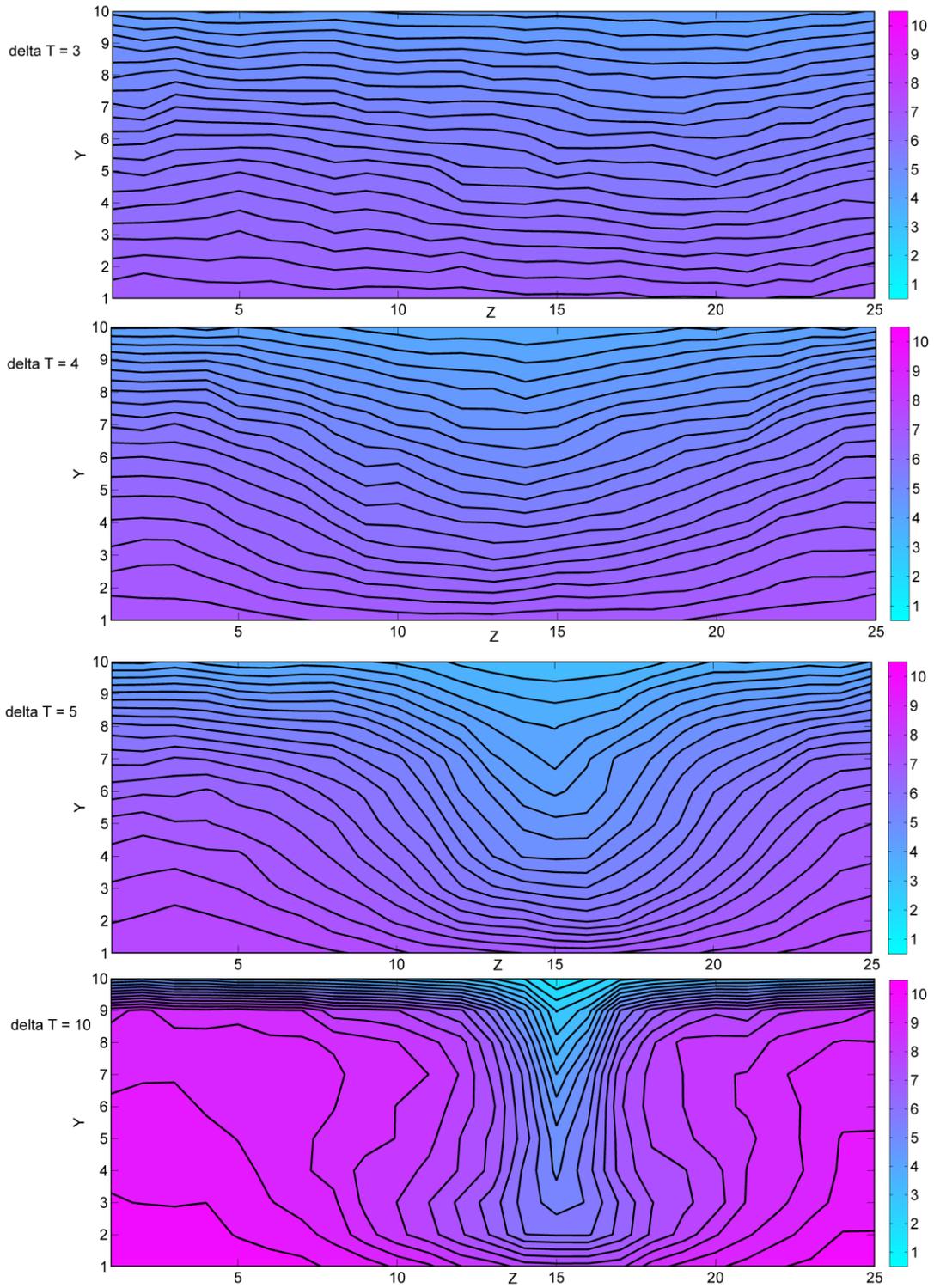

**Fig. 4**

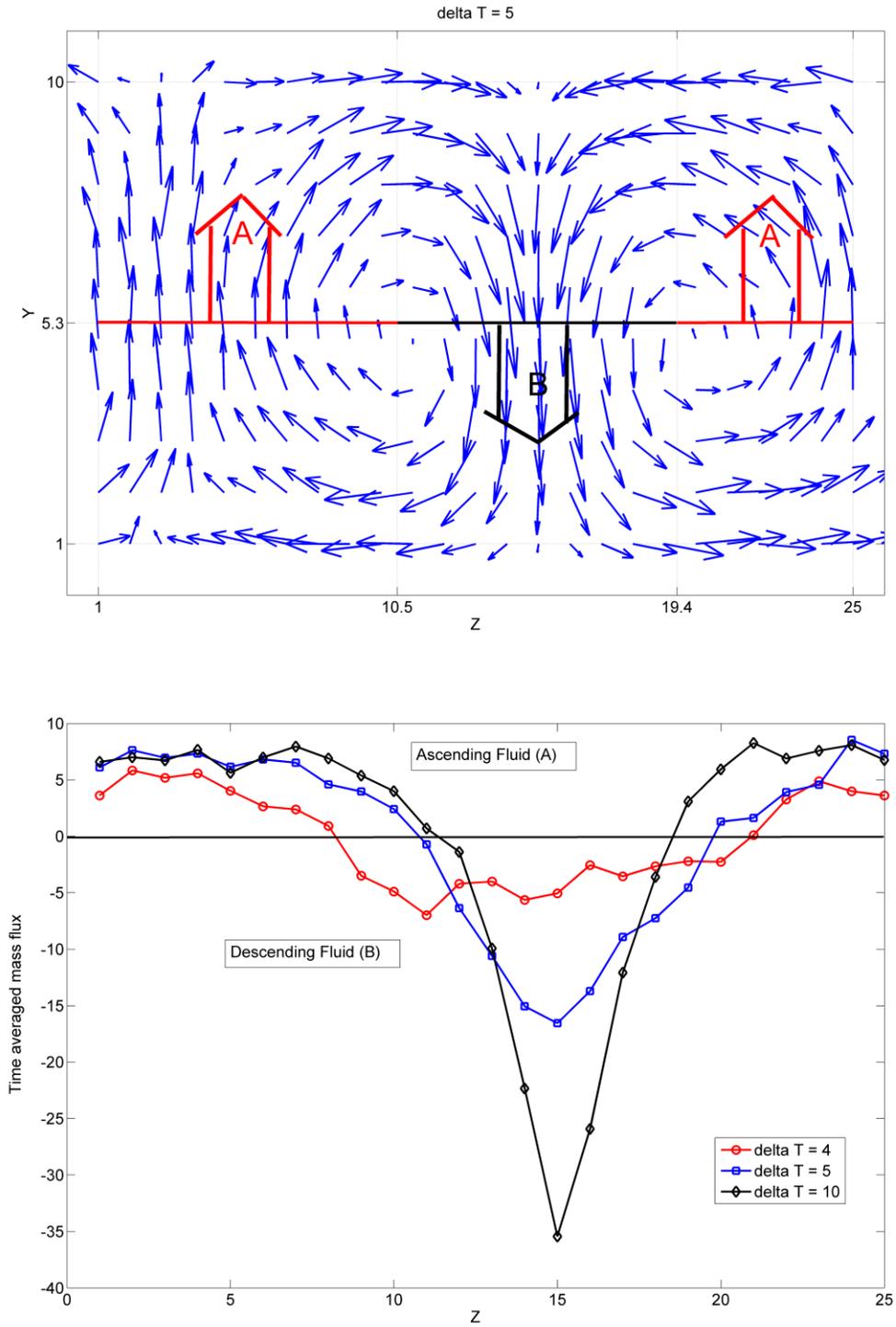

**Fig. 5**

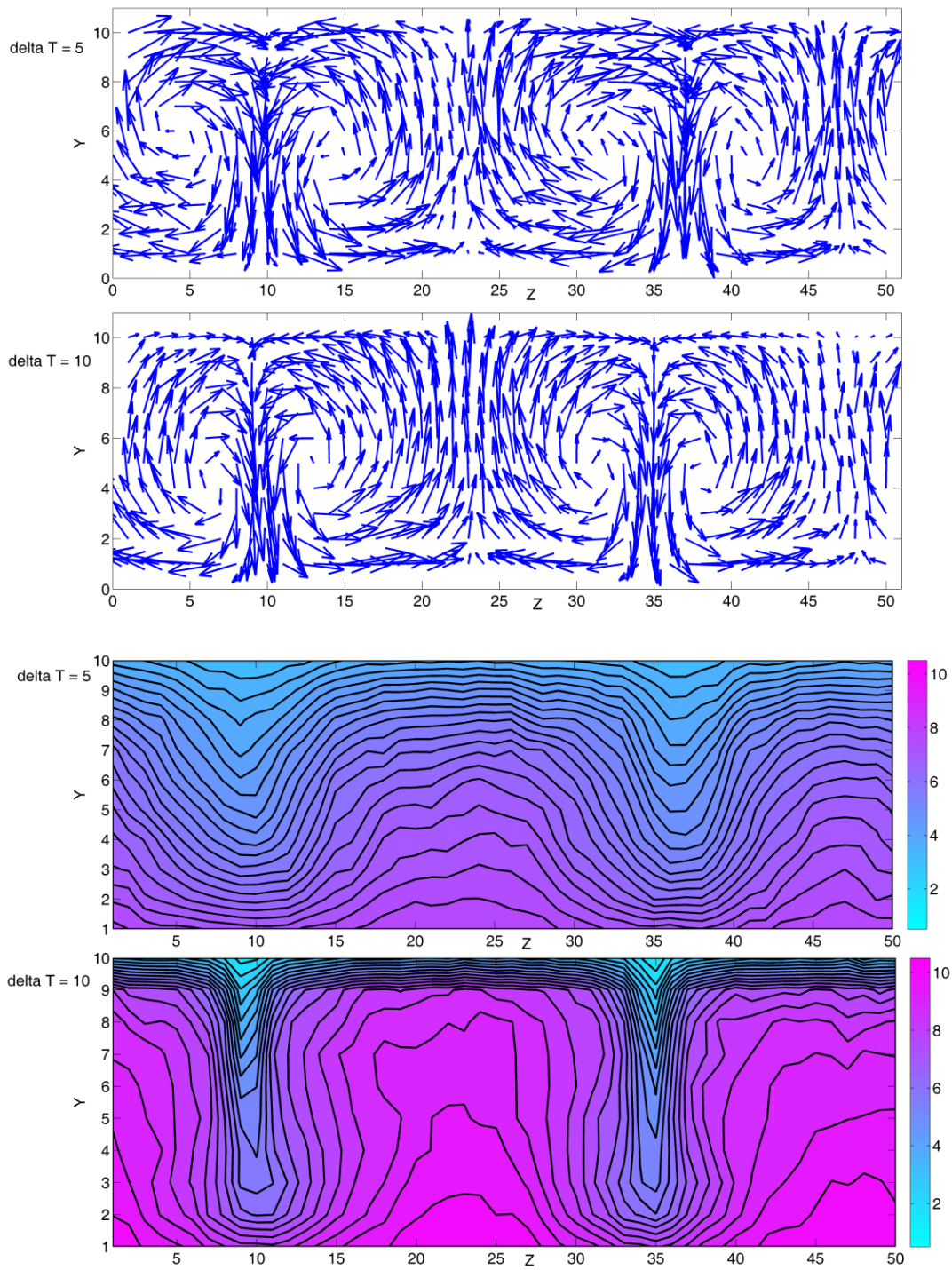

**Fig. 6**

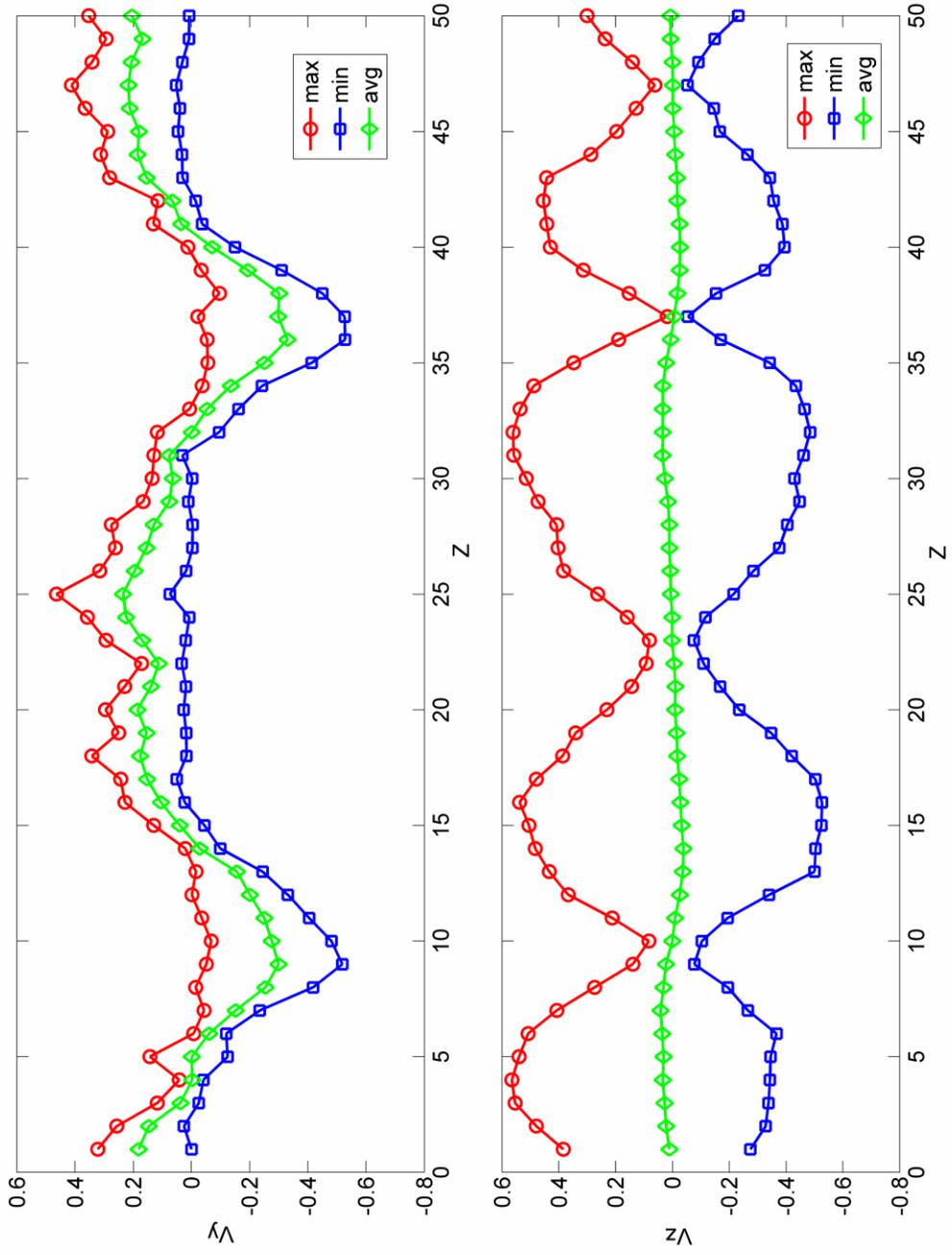

**Fig. 7**

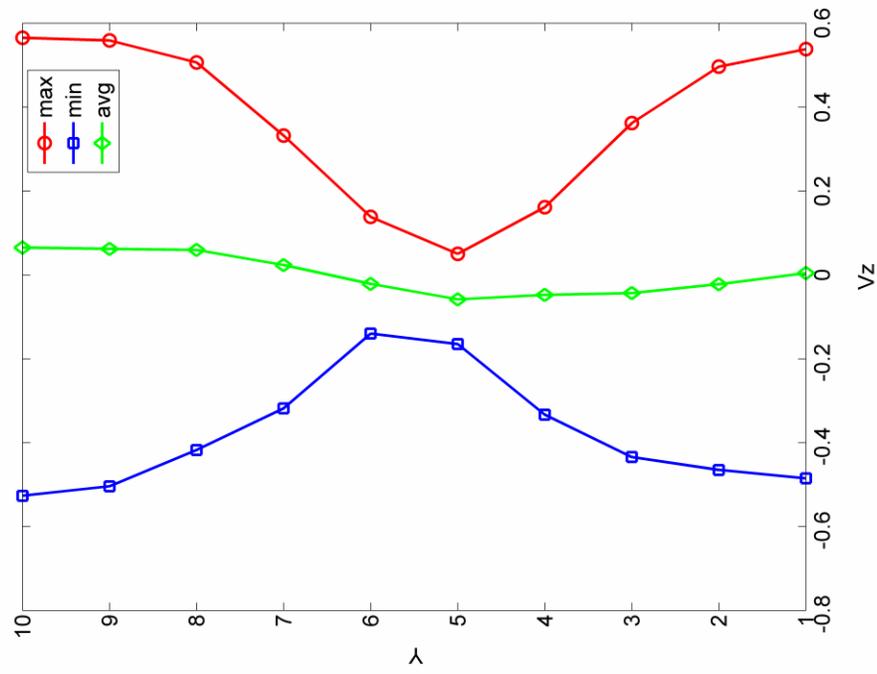
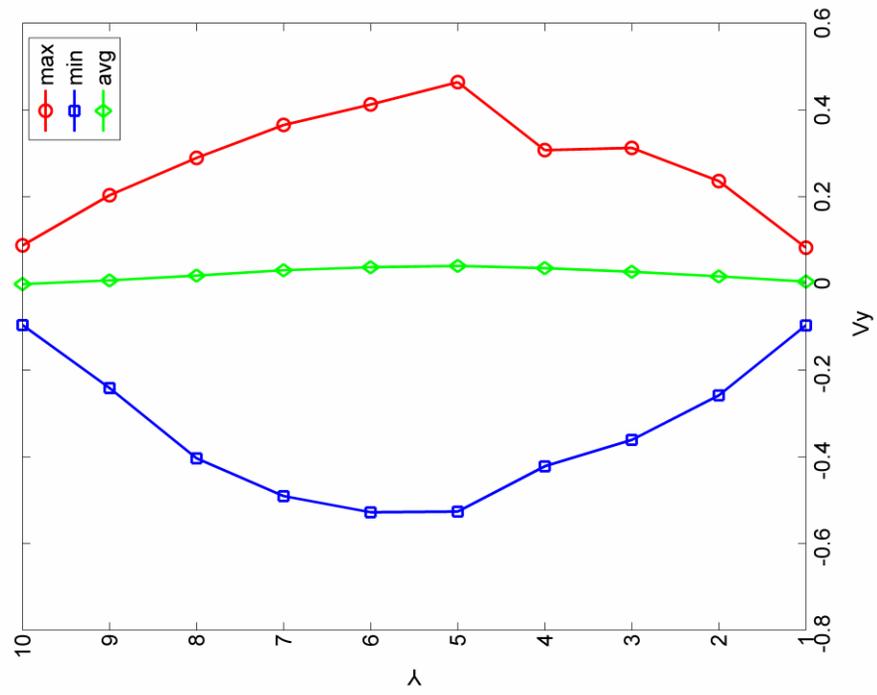

**Fig. 8**

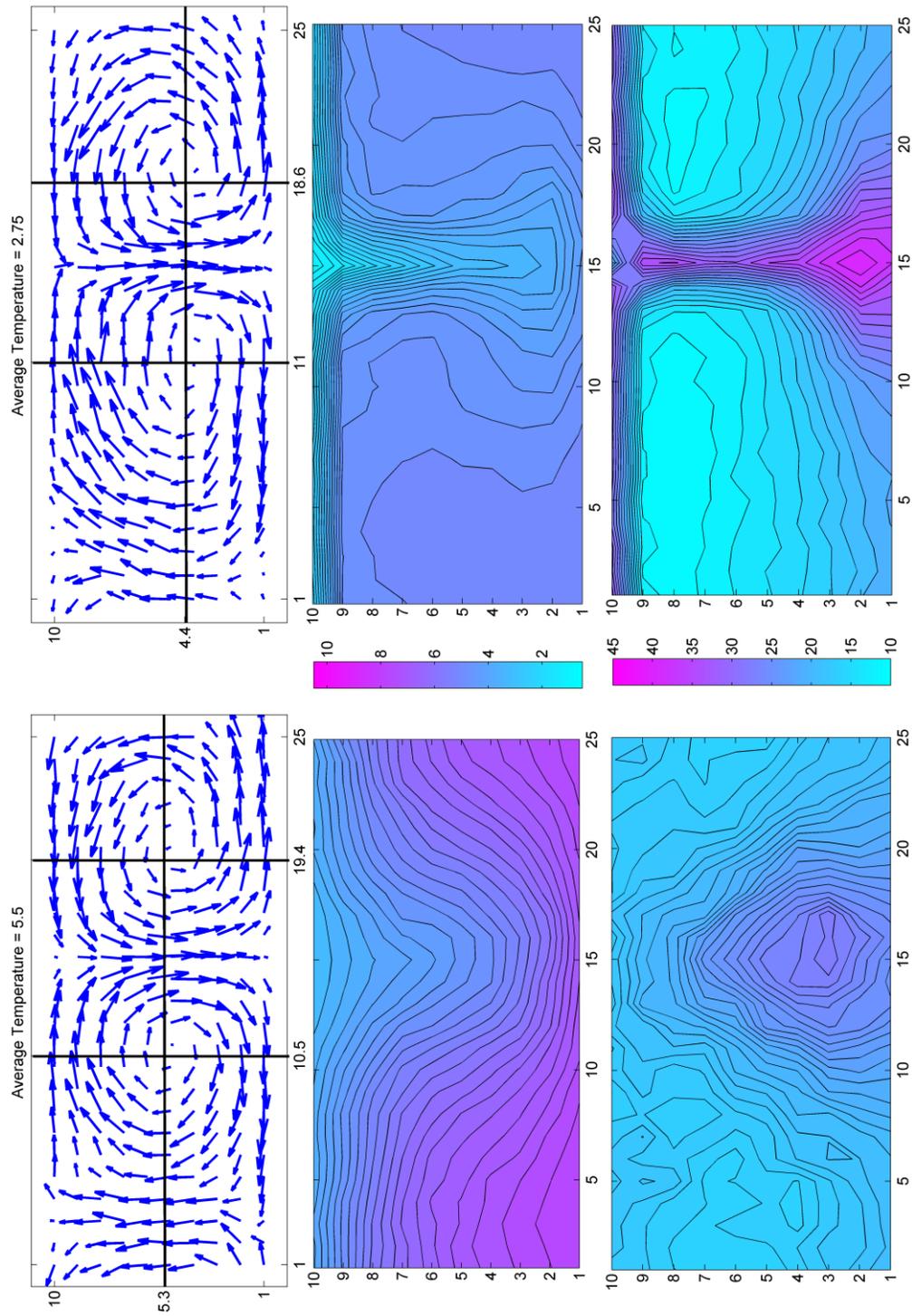

**Fig. 9**

**Table 1**

| PARAMETER | DESCRIPTION | VALUE |
|---|---|---|
| $\bar{r}_c$ | Interaction radius | 1.0 |
| $\bar{\rho}$ | Dimensionless density | 20.0 |
| $\bar{m}$ | Mass of bead | 1.0 |
| $\bar{\sigma}$ | Noise (random) parameter | 5.0 |
| $\bar{\gamma}$ | Friction (dissipative) parameter | $\bar{\gamma} = \frac{\bar{\sigma}^2}{4}\left[\frac{1}{\bar{T}_i} + \frac{1}{\bar{T}_j}\right]$ |
| $\bar{a}$ | Conservative force parameter | 0.0 |
| $\Delta \bar{t}$ | Time step | $5 \times 10^{-3}$ |
| $\lambda$ | Factor in GW integrator | 0.65 |
| $\bar{\varepsilon}_c$ | Heat capacity | $10^3$ |
| $\bar{K}_o$ | Strength of heat conduction | $10^{-8}$ |
| $\bar{K}^\varepsilon$ | Mesoscopic heat conductivity | $\bar{K}^\varepsilon = \bar{K}_o \frac{\overline{\varepsilon_i + \varepsilon_j}^2}{4}$ |
| $\bar{\Lambda}^\varepsilon$ | Random heat flux parameter | $\left[\bar{\Lambda}^\varepsilon\right]^2 = 2K^\varepsilon$ |
| $\bar{\varepsilon}_i$ | Bead internal energy | $\bar{\varepsilon}_i = \bar{\varepsilon}_c \bar{T}_i$ |
| $\bar{g}$ | Strength of external force | 0.5 |
| $\overline{k_B T_{\text{avg}}}$ | Average temperature of beads | See Table 2 |
| $N$ | Number of DPD beads | See Table 2 |
| $\bar{\Omega}$ | System volume | See Table 2 |
| $N_{\text{steps}}$ | Number of iterations | See Table 2 |

**Table 2**

| $N$ | $\bar{\Omega}$ | $N_{\text{steps}}$ | $\overline{k_B T_{\text{hot}}}$ | $\overline{k_B T_{\text{cold}}}$ | $\bar{\delta}$ | $\overline{k_B \Delta T}$ | $\dfrac{\overline{k_B \Delta T}}{\overline{k_B T_{\text{avg}}}}$ | $\bar{R} = \dfrac{\bar{m}\bar{g}\bar{H}}{\overline{k_B \Delta T}}$ |
|---|---|---|---|---|---|---|---|---|
| $\overline{k_B T_{\text{avg}}} = 5.5$ ||||||||||
| 5000 | 1x9x25 | 9 x 10⁵ | 7 | 4 | 2.78 | 3 | 0.5455 | 1.5 |
|  |  |  | 7.5 | 3.5 |  | 4 | 0.7273 | 1.125 |
|  |  |  | 8 | 3 |  | 5 | 0.9091 | 0.9 |
|  |  |  | 10.5 | 0.5 |  | 10 | 1.8182 | 0.45 |
| 10000 | 1x9x50 | 3 x 10⁵ | 7 | 4 | 5.56 | 3 | 0.5455 | 1.5 |
|  |  |  | 7.5 | 3.5 |  | 4 | 0.7273 | 1.125 |
|  |  |  | 8 | 3 |  | 5 | 0.9091 | 0.9 |
|  |  |  | 10.5 | 0.5 |  | 10 | 1.8182 | 0.45 |
| 7500 | 1x14x25 | 4 x 10⁵ | 6.5 | 4.5 | 1.79 | 2 | 0.3636 | 3.5 |
|  |  |  | 7 | 4 |  | 3 | 0.5455 | 2.33 |
|  |  |  | 7.5 | 3.5 |  | 4 | 0.7273 | 1.75 |
|  |  |  | 9.25 | 1.75 |  | 7.5 | 1.3636 | 0.93 |
| $\overline{k_B T_{\text{avg}}} = 2.75$ ||||||||||
| 5000 | 1x9x25 | 9 x 10⁵ | 5.25 | 0.25 | 2.78 | 5 | 1.8182 | 0.9 |

**Table 3**

| TEMPERATURE DIFFERENCE | MASS FLOW RATE OF ASCENDING FLUID (A) | MASS FLOW RATE OF DESCENDING FLUID (B) |
|---|---|---|
| 4 | 45.09 | 46.62 |
| 5 | 71.94 | 83.78 |
| 10 | 105.46 | 110.73 |